# Suppression of diamagnetism in superconducting nanorings by quantum fluctuations

TERHI T. HONGISO and KONSTANTIN YU. ARUTYUNOV*

*Nano Science Center, Department of Physics, University of Jyväskylä, PB 35, 40014, Jyväskylä, Finland*

e-mail: konstantin.arutyunov@phys.jyu.fi

**Zero resistance, diamagnetism (Meissner effect) and the energy gap in the excitation spectrum are the fundamental attributes of superconductivity. In nanoparticles with characteristic size ~ 10 nm superconductivity is suppressed by the quantization of the electron energy spectrum[1]. In narrow quasi-1D superconducting channels thermal fluctuations enable zero resistance only at temperatures noticeably below the critical temperature $T_c$ [2,3], while quantum fluctuations suppress the dissipationless electric current in ultra-narrow superconducting nanowires even at temperatures T→0 [4,5,6,7,8]. Here we demonstrate that the same essentially nanoscale phenomenon – *quantum fluctuations* – is responsible for quenching diamagnetism in superconducting nanorings dramatically affecting the magnitude, the period and the shape of the circulating persistent currents (PCs). The effect is of fundamental importance for our understanding of the behavior of a quantum coherent system at nanoscales; and should be taken in consideration in various nanoscience applications.**

It is a text-book knowledge that superconductivity is a collective phenomenon. Hence, one may ask a reasonable question: What should be the minimal dimension of a system to sustain the "collective" behavior? The answer depends on the particular system and the superconducting feature under discussion. In nanograins with diameter of about 10 nm [1] superconductivity is excluded when the size-dependent quantized energy level spacing δ is comparable to the superconducting gap $\Delta$[9]. Finite resistance at T<<$T_c$ is observed in superconducting nanowires with diameter ≤ 15 nm [4,5,6,7] enabling a certain rate of quantum fluctuations [8]. When a bulk superconductor is exposed to an external magnetic field the PCs at the surface repel the magnetic field contributing to the perfect diamagnetism. However, straightforward magnetization measurements of superconducting particles with dimensions smaller than the field penetration depth λ (for conventional superconductors of about 50 nm) provide extremely weak signal due to the negligible magnetic field repulsion from the bulk. Magnetization experiments on tin nanospheres down to ~80 nm[10]; indium, lead and aluminium nanoparticles down to ~ 10 nm[11,12] were not conclusive to state any size-dependent suppression of the PCs responsible for the diamagnetism.

The objective of this work was to study the size dependence of the PCs circulating in tiny superconducting rings using not magnetization, but extremely sensitive solid state tunneling technique. Application of an external magnetic field to a coherent non-single-connected system (e.g. superconducting ring) results in periodic oscillations of the PCs. The early experiments on relatively massive superconducting cylinders revealed the clear quantization of the magnetic momentum[13,14] with the period $\Delta\Phi$ equal to the flux quantum $\phi_0 = h/2e$. Later experiments on superconducting microrings at temperatures significantly below the critical one[15,16,17,18] revealed PC oscillations with much higher periods equal to the integer multiples of the flux quantum $\Delta\Phi / \phi_0 = n \sim 70$.

The allowed energy levels $E_n$ of a thin-walled superconducting ring in perpendicular magnetic field are given by the set of parabolas (Fig. 1a). If the system always relaxes to the ground state, the evolution of the energy in magnetic field is represented in Fig. 1a by the thick solid line resulting in periodicity strictly equal to one flux quantum: $\Delta\Phi = \phi_0$. That is the typical case in experiments at temperatures not much below the critical one[13,14,19] or in systems containing weak links – e.g. SQUIDs. The corresponding dependence of the PCs on magnetic flux $I_s \sim dE_n/d\Phi$ follows the well-known saw-tooth pattern with the single flux quantum periodicity (Fig. 1b, thick solid line). However, at temperatures $T \ll T_c$ the system can be "frozen" in a metastable state n without relaxing to the neighboring quantum level $n \pm 1$. The ultimate requirement for the transition to another quantum state is that the circulating PC reaches its critical value $I_c$ (Fig. 1b, horizontal line). It can be shown[16,20] that the corresponding period of PC oscillations $\Delta\Phi / \phi_0 \sim S / \xi$, where $\xi$ is the superconducting coherence length. Hence, if (i) the temperature is low enough to enable metastable state formation, and (ii) the circumference of the loop is large $S \gg \xi$ one should observe periodicity $\Delta\Phi / \phi_0 \sim S / \xi \gg 1$ in a full accordance with the experiments[15,16,17,18].

With the rapid development of nanotechnology it was found [4,5,6,7,8] that in superconducting nanowires with the effective diameter $\leq 15$ nm the quantum fluctuations of the order parameter, also called *quantum phase slips*, suppress dissipationless supercurrent leading to experimentally observable finite resistance at temperatures $T \ll T_c$. When a thin-walled superconducting nanoring is exposed to an external magnetic field the same effect – quantum fluctuations – should lead to a qualitatively new effect[21]: the gap opens in the energy spectrum and the PC saw-tooth pattern degenerates into a smooth sine-type dependence (Fig. 1, dotted lines). So far theoretical prediction[21] has not been confirmed experimentally. Remarkably, for the given wire cross section forming the ring the magnitude of the PC oscillations depends exponentially on the diameter of the loop $I_{PC} \sim \exp(-R/R_c)$ with the critical radius $R_c$ of few micrometers[21,8]. Recovering the above discussion about the high-periodic oscillations of PC, one can predict that in a superconducting nanorings with fixed circumference $\xi << S \sim 2\pi R_c$ the reduction of the wire cross section should result not only in changes of the magnitude and the shape of the PC oscillations[21], but also their *period* should drop from values $\Delta\Phi / \phi_0 \sim S / \xi >> 1$ down to $\Delta\Phi / \phi_0 = 1$.

To test these hypotheses we fabricated all-aluminium superconducting tunnel structures with the central electrode in the shape of a loop (Fig. 2) and utilized the tunnel spectroscopy method to probe the PCs [18,20]. In zero magnetic field all samples demonstrated conventional I-V dependencies (Fig. 3, inset). Then the bias voltage was fixed below the quadruple energy gap $V \leq 4\Delta/e$ and the perpendicular magnetic field was applied. In the originally "thick" structures the tunnel current $I_T$ showed pronounced saw-tooth oscillations (Figs. 3 and 4) with the "large" period $\Delta\Phi / \phi_0 > 1$ in a full agreement with the earlier findings[18,20]. We used ion beam etching to progressively and non-destructively reduce the cross section $\sigma$ of the wire forming the loop[22,23]. The approach enables tracing the evolution of a size phenomenon on a same nanostructure without introducing uncertainties coming from uniqueness of samples fabricated in different technological runs. No degradation of the shape of the I-V characteristics in zero field was observed with the reduction of the wire cross section (Fig. 3, inset). Slight increase of the tunnel resistance at $eV > 4\Delta$ can be attributed to shrinking of the tunnel contact area while sputtering or/and ion beam stimulated annealing of the junction itself [22]. Contrary to the I-Vs, the pattern of the PC oscillations in magnetic field exhibit dramatic changes with reduction of the cross section $\sigma$ of the wire forming the loop. First, the magnitude of the oscillations at a given voltage bias $I_T (\Phi, V=const)$ goes down (Fig. 3). Second, the period of the oscillations decreases. For example, in the samples with the loop circumference S= 19 μm the period gradually drops from $\Delta\Phi / \phi_0 = 6$ down to 1 (Fig. 4). And, finally, the shape of the oscillations changes: from pronounced saw-tooth-type to quasi-sine-type (Fig. 4). With further reduction of the wire cross section $\sigma$ any traces of the $I_T(\Phi)$ oscillations disappear: either the loop trivially breaks, or the quantum fluctuations suppress the magnitude of the PC oscillations down to the level making experimental observation impossible. Qualitatively similar results were obtained on several structures with two different sizes of the loop.

All structures were subjected to an extensive AFM/SEM analysis. Only those samples which contained no obvious structural imperfections or other sorts of weak links were processed. According to our earlier experiments with ultra-narrow aluminium nanowires with diameters down to ~ 8 nm[6,7,23], the low energetic $Ar^+$ ion beam sputtering provides smoothening of the surface "healing" the inevitable imperfections. It is very improbable that the ion beam treatment can introduce structural inhomogeneities or reveal initially hidden ones, which cannot be detected by SEM or AFM. In the studied structures with the effective wire diameters from ~30 nm to ~70 nm the existence of short constrictions (notches) can be ruled out completely (Figs. 2b and 2c). Additionally, the unique property of aluminium nanowires - the increase of the critical temperature with reduction of the diameter[6,7,24,25] - makes the hypothetical constrictions not effective weak links. Similar results obtained on several sets of samples do not favor the "constriction" scenario, which should be a unique property of each individual structure. Particularly, the gradual reduction of the $I_T(\Phi)$ oscillation period with the virtually constant base line (Figs. 3 and 4) would be difficult to explain. Resuming, with a high level of confidence we may state that the trivial explanation of our observations by formation of a SQUID-type structure (loop with a weak link) is not credible. On the contrary, the depicted quantum fluctuation scenario[21,8] qualitatively explains our findings. As the probability of the quantum phase slip exponentially depends on the wire cross section $\sigma$[8], in real (not perfectly uniform) structures the dominating contribution comes from the thinnest parts[24,7]. The observation can complicate the quantitative comparison with theory[21,8], but cannot change the main conclusion of the work: in superconducting rings the quantum fluctuations suppress the metastable quantum states gradually reducing the magnitude and the period of PC oscillations.

The quenching of PCs in ultra-narrow rings is yet another manifestation of the quantum fluctuations in superconductors. Additionally to the already observed quenching of superconductivity in small grains[1,9] and the destruction of the zero resistance state in ultra-narrow 1D channels[4,5,6,7], in this work we show that the quantum fluctuations are responsible for suppression of diamagnetism in superconducting nano-sized systems. Apart from the basic science importance, the discovery is crucial for numerous applications setting the fundamental limitations on miniaturization of superconducting components. Besides this pessimistic consequence, utilization of the quantum fluctuation phenomenon is expected to result in building of a new class of devices: quantum standard of electric current[26] and q-bit[27].

**Methods.**

Lift-off electron beam lithography (EBL) technique was used to fabricate the tunnel structures. The samples consist of the ellipse shaped ring (Al) contacted by two electrodes (Al) on the opposite sides of the loop through tunnel junctions (Fig. 2). The oval shape of the central electrode was selected to eliminate parasitic overlapping shadows emerging in two-angle metal evaporation process. The tunnel barriers between the layers was created by controlled oxidation of the first 80 nm thick aluminium layer (the contacts) before evaporating the second 85 nm thick layer (the loop).

The samples were studied as a function of the cross section of the wire forming the ring. The reduction of the cross section was realized by the low energetic ion beam sputtering. Utilization of $Ar^+$ ions at acceleration voltages $\leq 1$ keV can be considered as virtually introducing no defects[22,23] with the ion penetration depth of the order of the thickness of the naturally grown oxide (about 3 nm in Al). Additionally, the sputtering provides a polishing effect healing the inevitable surface imperfections of the lift-off fabricated nanostructures[23].

The dimensions of the fabricated samples were measured with atomic force (AFM) and scanning electron (SEM) microscopes. However the AFM/SEM study after each sputtering would pose a too high threat of damaging the extremely fragile structures. In practice, the AFM/SEM measurements were taken after the first (high dose) sputtering step and finally at the end of the last measuring session (typically when the structure is already damaged). The sample dimensions at the intermediate steps were defined by interpolation of the known initial and final AFM/SEM data. The method results in high relative errors in determination of the cross section of the line forming the loop. However, according to our previous studies of aluminium nanowires[6,7], enabling independent determination of their cross section by measurement of the resistance, after multiple ion beam sputterings the typical surface inhomogeneity is about ± 3 nm. Hence, the large errors in Figs. 3 and 4 should be considered as the uncertainty in defining the effective mean value, and not the variation of the line cross section for the given sample.

After each step of reducing the cross section the samples were cooled down to temperatures below 100 mK using $He^3/He^4$ dilution refrigerator. The measurements were performed in an electromagnetically shielded environment using battery powered front-end amplifiers. The tunnel current over the whole structure was measured as a function of the perpendicular magnetic field intensity at several chosen constant bias voltages (Fig. 2a). The magnetic field sweeps (covering typically 5 to 10 periods of the oscillations) ranged from few minutes up to several hours. At temperatures below ~500 mK the data is quantitatively indistinguishable. Time-dependent 'jumps' of the tunnel current at a fixed voltage bias and fixed magnetic field are observed only above that temperature. The majority of the presented data has been collected at T≈50 mK, where no sweep rate dependent effects were observed

The effective area of the loop electrode calculated from the periodicity was found to vary from a cool down to another one less than 0.5 %, which is of the same order as within a single measurement session. The difference between the calculated effective area and the one measured from SEM/AFM analysis was found to be also about 0.5 %. The error in defining the magnetic field at the sample due to misalignment of the coil was estimated to be about 0.05 %, hence being smaller than the above indicated uncertainties.

**Author contributions**

T.T.H. designed the structures, performed experiments and analyzed the data. K.Y.A. planned the project and interpreted the results.

**Acknowledgements**

Authors would like to acknowledge F. Hekking, T. Klapwijk, D. Vodolazov and A. Zaikin for valuable discussions; M. Zgirski and P. Jalkanen for assistance with the ion beam treatment; and J. Lehtinen for help with the sample fabrication; and A. Julukian for the help with AFM analysis. The work was supported by the Finnish Academy project FUNANO.

No **Competing Financial Interests** are involved.

**Figure legends.**

Fig. 1. **Dependence of the energy E and persistent current $I_S$ on magnetic flux.** (a) Quantized energy spectrum $E_n$ of a thin-walled superconducting ring in external magnetic flux $\Phi$ normalized by the flux quantum $\phi_0=h/2e$. Solid line corresponds to the evolution of the system following the ground state, dashed line – metastable states, and the dotted line – energy levels in the limit of strong quantum fluctuations. (b) Corresponding dependence of the persistent current $I_s$. Horizontal dashed line corresponds to the critical current $I_c$ setting the ultimate limit for the magnitude of the persistent currents circulating in the loop.

Fig. 2. M**icroscope images of the structures**. (a) SEM image of the double junction all-aluminium SISIS tunnel structure; and schematics of the electric measurements. (b) AFM image of a typical structure just after fabrication. (c) AFM image of the same structure after ion treatment. One can clearly see the reduction of the cross section of the wire forming the loop and polishing of the surface. The histograms analyzing the distribution of the cross sections of the loops are plotted in the insets. Note that in both cases the lowest cross section is well-defined and is finite indicating the absence of narrow constrictions.

Fig. 3. **Oscillations of the tunnel current $I_T$ in external magnetic flux $\Phi$ normalized by the flux quantum $\phi_0=h/2e$.** Variation of the tunnel current (~ persistent current in the ring) in magnetic field measured on the same nanostructure with progressively reduced cross section $\sigma$ of the wire forming the loop. One can clearly see both the reduction of the magnitude and the period of the oscillations. Arrows indicate the directions of the magnetic field sweeps. The long-period oscillation at the beginning of the magnetic field sweep corresponds to the evolution of the system within the energy level with the same quantum number. Magnetic field sweep time was about half an hour. Inset: zero field I-V characteristics of the samples depicted in the main panel. Arrow indicate the voltage bias V=0.608 mV at which the magnetic field sweeps were taken. Measurements were made at T = 62 ± 5 mK and T = 54 ± 5 mK for the bottom and the top I-V's referred to the V = 1 mV point.

Fig. 4. **Evolution of the tunnel current $I_T$ oscillations in external magnetic flux $\Phi$ normalized by the flux quantum $\phi_0$=h/2e.** Tunnel current oscillations measured on samples with the same circumference of the loop S =19 μm but with different effective diameters of the wire forming the loop $\sigma^{1/2}$ : (□) 70 ± 35 nm, T = 93 ± 5 mK; (○) 55 ± 27 nm, T=93 ± 5 mK ;(Δ) 60 ± 33 nm, T = 62 ± 5 mK; (●) 35 ± 25 nm, T=54 ± 5 mK. For details in determination of the wire cross section σ see the "Methods". One can clearly see the reduction of the period of oscillations and the degeneration of the saw-tooth shape into the sine-type pattern for the thinnest sample. Vertical bars indicate the scales. As the magnetic field sweeps were taken at slightly different bias points, the reduction of the magnitude of the oscillations is less obvious compared to Fig. 3. Arrows show the directions of the magnetic field sweeps each taken on a time scale about half an hour.

# References

[1] Black, C. T., Ralph, D. C. & Tinkham, M. Spectroscopy of the superconducting gap in individual nanometer-scale aluminum particles. *Phys. Rev. Lett.* **76**, 688-689 (1996).

[2] Langer, J. S. & Ambegaokar, V. Intrinsic resistive transition in narrow superconducting channels. *Phys. Rev.* **164,** 498-510 (1967).

[3] McCumber, D. E. & Halperin, B. I. Time scale of intrinsic resistive fluctuations in thin superconducting Wires. *Phys. Rev. B* **1,** 1054-1070 (1970).

[4] Giordano, N. Evidence for macroscopic quantum tunneling in one-dimensional superconductors. *Phys. Rev. Lett.* **61,** 2137-2140 (1988).

[5] Bezryadin, A., Lau, C. N. & Tinkham, M. Quantum suppression of superconductivity in ultrathin nanowires. *Nature* **404,** 971-974 (2000).

[6] Zgirski, M., Riikonen, K.-P., Touboltsev, V. & Arutyunov, K. Size dependent breakdown of superconductivity in ultranarrow nanowires. *Nano Lett.* **5,** 1029-1033 (2005).

[7] Zgirski, M., Riikonen, K.-P., Touboltsev, V. & Arutyunov, K. Yu. Quantum fluctuations in ultranarrow superconducting aluminum nanowires. *Phys. Rev. B* **77,** 054508 (2008).

[8] Arutyunov, K. Yu., Golubev, D. S. & Zaikin, A. D. Superconductivity in one dimension. *Phys. Rep.* **464,** 1-70 (2008).

[9] von Delft, J., Zaikin, A.D., Golubev, D.S. & Tichy, W. Parity-affected superconductivity in ultrasmall metallic grains. *Phys. Rev. Lett.* **77**, 3189-3192 (1996).

[10] Naumenko, I. G., Petinov, V. I. & Gen, M. Ya. High-frequency magnetic polarizability of small superconducting tin particles, *Sov. Phys. – Solid State*, **13**, 2740-2744 (1972).

[11] Matsuo, S., Sugiura, H. & Noguchi, S. Superconducting Transition Temperature of Aluminum, Indium, and Lead Fine Particles, *J. Low. Temp. Phys.*, **15**, 481-490 (1974).

[12] Tsai, A. P., Chandrasekhar, N. & Chattopadyay, K. Size effect on the superconducting transition of embedded lead particles in an Al-Cu-V amorphous matrix. *Appl. Phys. Lett.* **75**, 1527-1528 (1999).

[13] Doll, R. & Näbauer, M. Experimental proof of magnetic flux quantization in a superconducting ring. *Phys. Rev. Lett.* **7,** 51-52 (1961).

[14] Deaver, B. S. & Fairbank, W. M. Experimental evidence for quantized flux in superconducting cylinders. *Phys. Rev. Lett.* **7,** 43-46 (1961).

[15] Pedersen, S., Kofod, G. R., Hollingbery, J. C., Sørensen, C. B. & Lindelof, P. E. Dilation of the giant vortex state in a mesoscopic superconducting loop. *Phys. Rev. B* **64,** 104522 (2001).

[16] Vodolazov, D. Y., Peeters, F. M., Dubonos, S. V. & Geim, A. K. Multiple flux jumps and irreversible behavior of thin Al superconducting rings. *Phys. Rev. B* **67,** 054506 (2003).

[17] Bourgeois, O., Skipetrov, S. E., Ong, F. & Chaussy, J. Attojoule Calorimetry of Mesoscopic Superconducting Loops. *Phys. Rev. Lett.* **94,** 057007 (2005).

[18] Arutyunov, K. Yu. & Hongisto, T.T. Normal-metal-superconductor interferometer. *Phys. Rev. B* **70**, 064514 (2004).

[19] Little, W. A. & Parks, R. D. Observation of quantum periodicity in the transition temperature of a superconducting cylinder. *Phys. Rev. Lett.* **9,** 9-12 (1962).

[20] Vodolazov, D. Y., Peeters, F. M., Hongisto, T.T. & Arutyunov, K. Yu. Microscopic model for multiple flux transitions in mesoscopic superconducting loops. *Europhys. Lett.* **75 (2),** 315-320 (2006).

[21] Matveev, K. A., Larkin, A. I. & Glazman, L. I. Persistent current in superconducting nanorings. *Phys. Rev. Lett.* **89,** 096802 (2002).

[22] Savolainen, M., Touboltsev, V., Koppinen, P., Riikonen, K.-P. & Arutyunov, K. Ion beam sputtering for progressive reduction of nanostructures dimensions. *Appl. Phys. A* **79,** 1769-1773 (2004).

[23] Zgirski, M., Riikonen, K.-P., Tuboltsev, V., Jalkanen, P., Hongisto, T. T. & Arutyunov, K. Yu. Ion beam shaping and downsizing of nanostructures. *Nanotechnology* **19,** 055301 (2008).

[24] Zgirski, M. & Arutyunov, K.Yu. Experimental limits of the observation of thermally activated phase-slip mechanism in superconducting nanowires. *Phys. Rev. B* **75,** 172509 (2007).

[25] Shanenko, A.A., Croitoru, M. D., Zgirski, M., Peeters, F. M. & Arutyunov, K. Size-dependent enhancement of superconductivity in Al and Sn nanowires: Shape-resonance effect. *Phys. Rev. B* **74,** 052502 (2006).

[26] Mooij, J. E. & Nazarov, Yu. V. Superconducting nanowires as quantum phase-slip junctions. *Nature Physics* **2,** 169, (2006).

[27] Mooij, J. E. & Harmans, C. J. P. M. Phase-slip flux qubits. *New J. Phys.* **7,** 219 (2005).

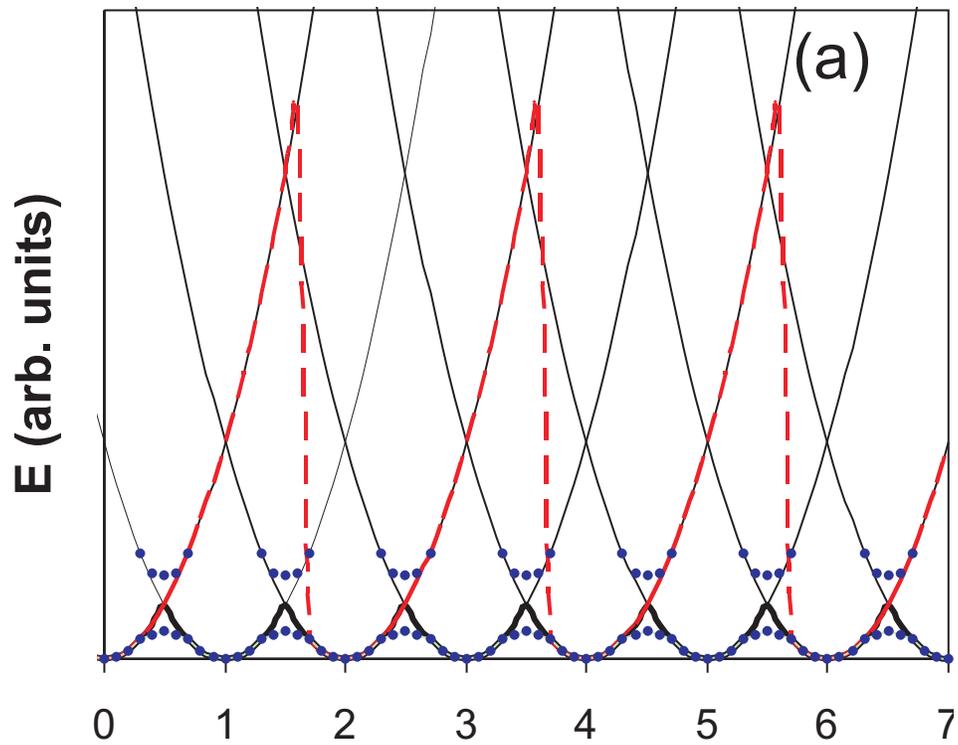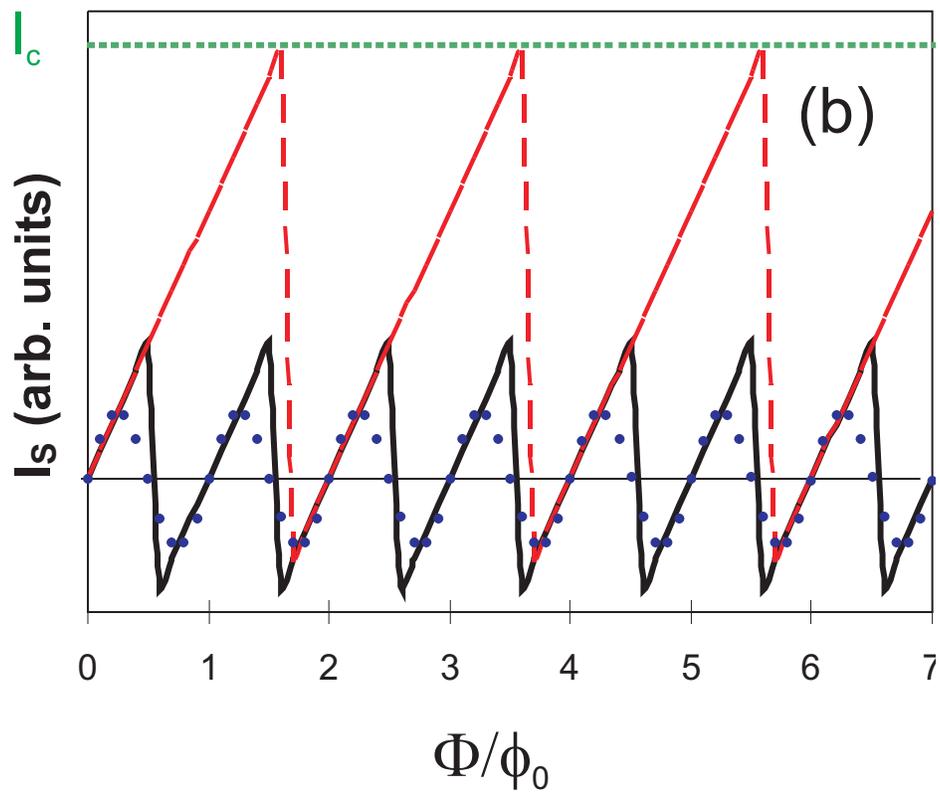

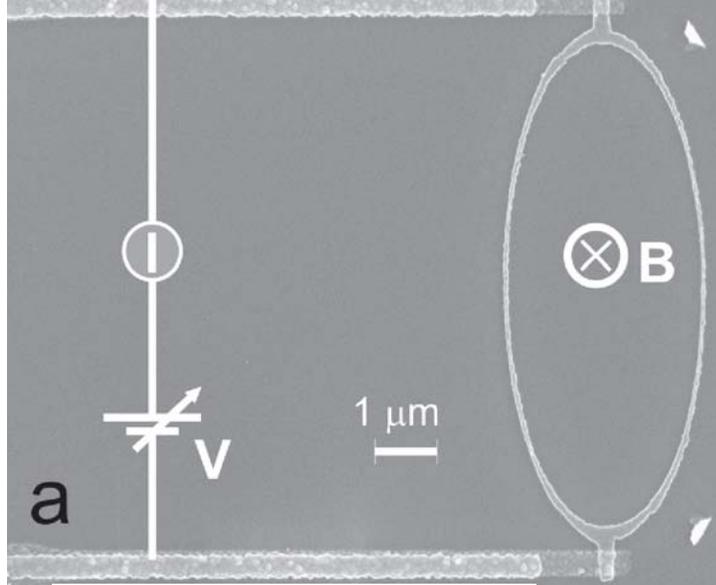

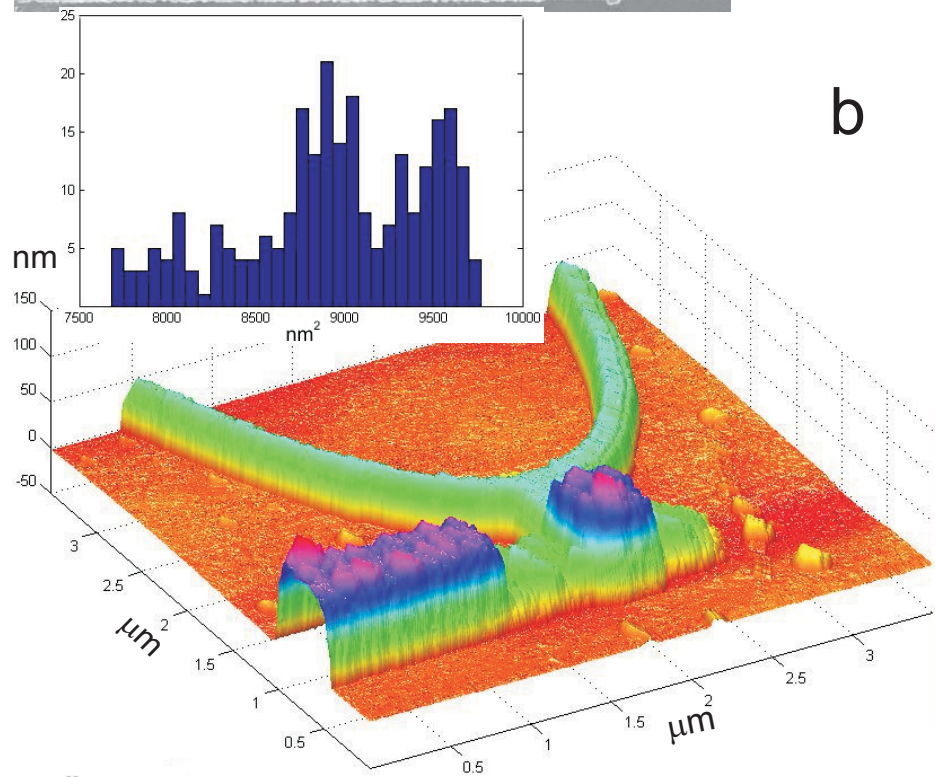

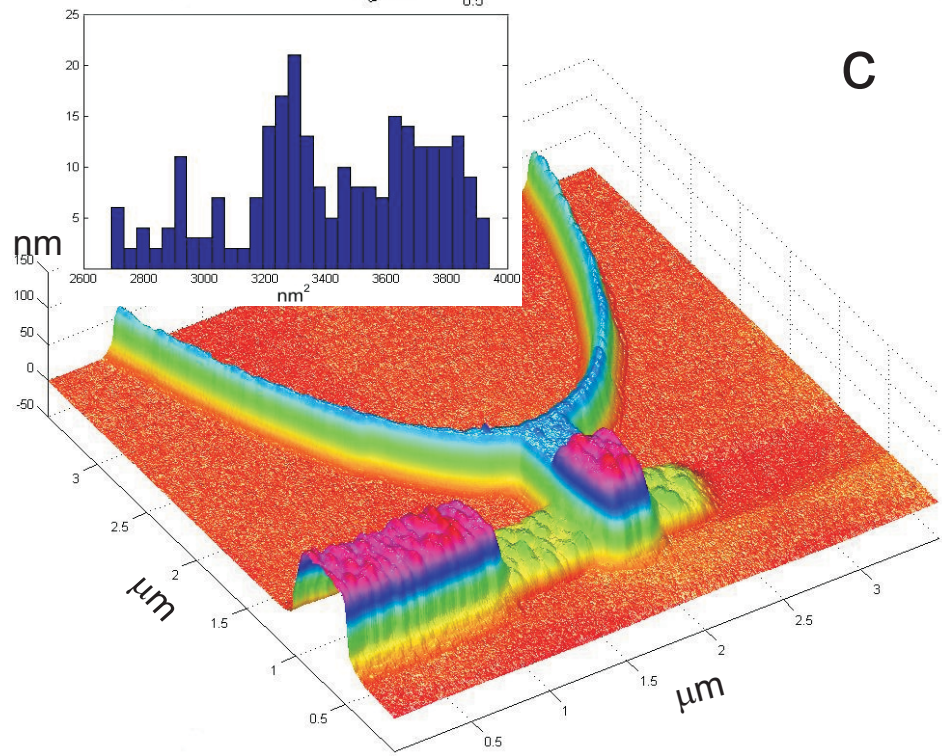

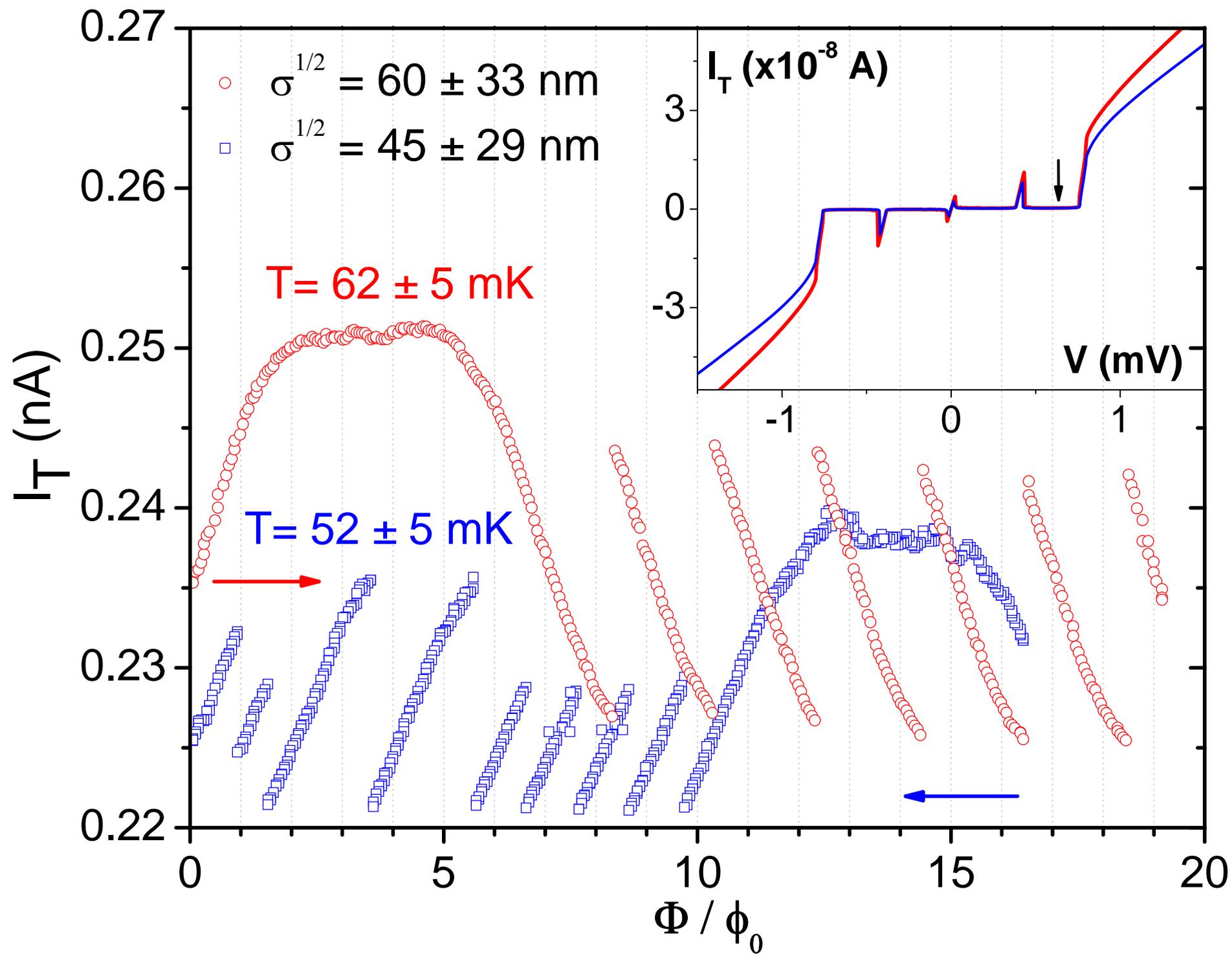

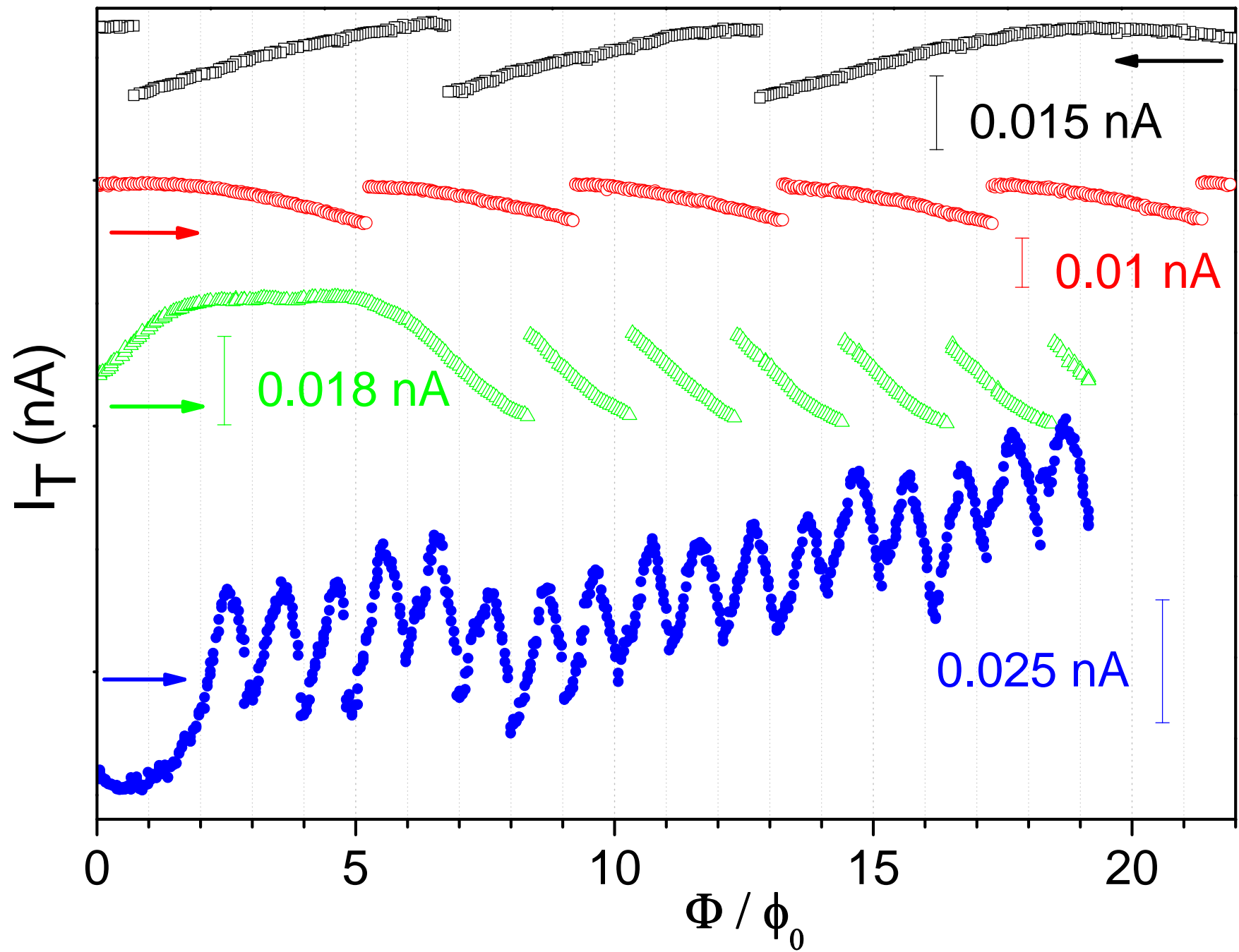

# Supplementary information.

Probability of quantum fluctuations in a quasi-1D channel exponentially depends on the cross section σ [8]. Hence, the homogeneity of studied structures is a crucial point in interpretation of the experimental data. Existence of inevitable geometrical imperfections in real samples should favor quantum phase slippage in these "weaker" sections. This observation complicates quantitative comparison with theory, while should not alter the main message – quantum fluctuations do govern the behavior of quasi-1D superconducting systems. In case of resistance measurements of the ultra-narrow nanowires [4-7], reasonable agreement with theory can be achieved just by substitution in the corresponding expressions the effective sample cross section, slightly smaller than the average geometric one.

In case of nanorings studied in the present work, existence of hypothetical weak links in the loop brings additional complicity to the interpretation of the experiment. If the weak link is "weak" enough, then it can act as a Josephson junction. If the case, the system can be considered as a SQUID-type structure giving rise to the corresponding periodicity $\Delta\Phi/\phi_0=1$ being a well-known effect. To test the behavior of such a system, in the same technological run we fabricated two structures, similar to the studied in the main part of the text: one with homogeneous ring-shaped central electrode, another – with the same dimensions, but containing a short and narrow constriction (notch) in the loop. The ratio of the line width in the constriction and in the rest of the loop was about 1:3. Oscillations of the tunnel current in magnetic field were studied in both structures. As expected, the homogeneous sample showed the high-periodic ($\Delta\Phi/\phi_0=6$) saw-tooth oscillations with the almost constant base line (Supplementary Fig-1). While in the structure with the notch noisy $\Delta\Phi/\phi_0=1$ periodicity superimposed on the non-monotonous envelope dependence with no traces of a regular saw-tooth $\Delta\Phi/\phi_0=6/3=2$ oscillations were detected (Supplementary Fig-2). We cannot exclude completely that the shape and the periodicity of oscillations in this notch-containing structure is also governed by the quantum fluctuations. However, compared to the "regular" saw-tooth pattern in the homogeneous sample, the noisy and non-monotonous $I_T(\Phi)$ dependencies in the second sample suggest that other phenomena

(e.g. Josephson effect and the corresponding SQUID-type physics) might contribute to the behavior of the system with the constriction.

Coming back to the data presented in the main part of the paper, we can conclude that the *gradual* reduction of the magnitude and the period of $I_T(\Phi)$ oscillations with the reduction of the wire cross section (Figs. 3 and 4) cannot be explained by the Josephson effect in the hypothetical weak link(s) of the loop-shaped electrode. As it has been clearly written in the main text, only the structures containing no obvious geometrical imperfections were studied. According to our previous experiments with aluminium nanowires down to diameters ~ 8 nm [6,7,24], due to the re-deposition of the sputtered material, short constrictions (and bumps) are effectively removed by the low energetic ion beam treatment. Only extended (and "smooth") variations of the line cross section remain. The unique property of aluminium nanostructures – increase of the critical temperature with reduction of the characteristic dimension [6,7,24,25] – makes these thinner parts not effective weak links. The segments of the loop with smaller cross section may serve as regions with higher probability of the phase slippage, but not as Josephson-type weak links.

Experiments with real (= not ideal) samples always contain an option for a critically-oriented reader to assign any effect to hypothetical imperfections (weak links, contact phenomena, boundary, etc.) and the corresponding "well-known" physics. In this work we paid special attention to the uniformity of the samples, and the persistent currents were probed using virtually non-invasive tunnel contacts. Combined with the large statistics obtained on multiple samples, we can state that the evolution of the shape, the period and the magnitude of the current oscillations with reduction of the wire cross section cannot be explained by a trivial formation of weak links in the loop electrode. On the contrary, the quantum phase slip scenario [21] nicely explains our findings.

## Supplementary figure legends.

**Supplementary figure-1.**

**Tunnel current oscillations in the structure with the uniform loop.**

Tunnel current oscillations $I_T(\Phi/\phi_0)$ in the SISIS tunnel structure with the uniform ring-shaped central electrode with diameter 5 μm at two closed voltage bias points. Measurements were made at T = 64 ± 5 mK. Arrows indicate the directions of the magnetic field sweeps taken at two closed voltage bias points.

**Supplementary figure-2.**

**Tunnel current oscillations in a structure with a loop containing constriction.**

Tunnel current oscillations $I_T(\Phi/\phi_0)$ in the structure of the same dimensions as in Supplementary Figure-1, but the central ring electrode contains short constriction (notch). Measurement was made at T= 72 ± 5 mK. Arrow indicates direction of the magnetic field sweep.

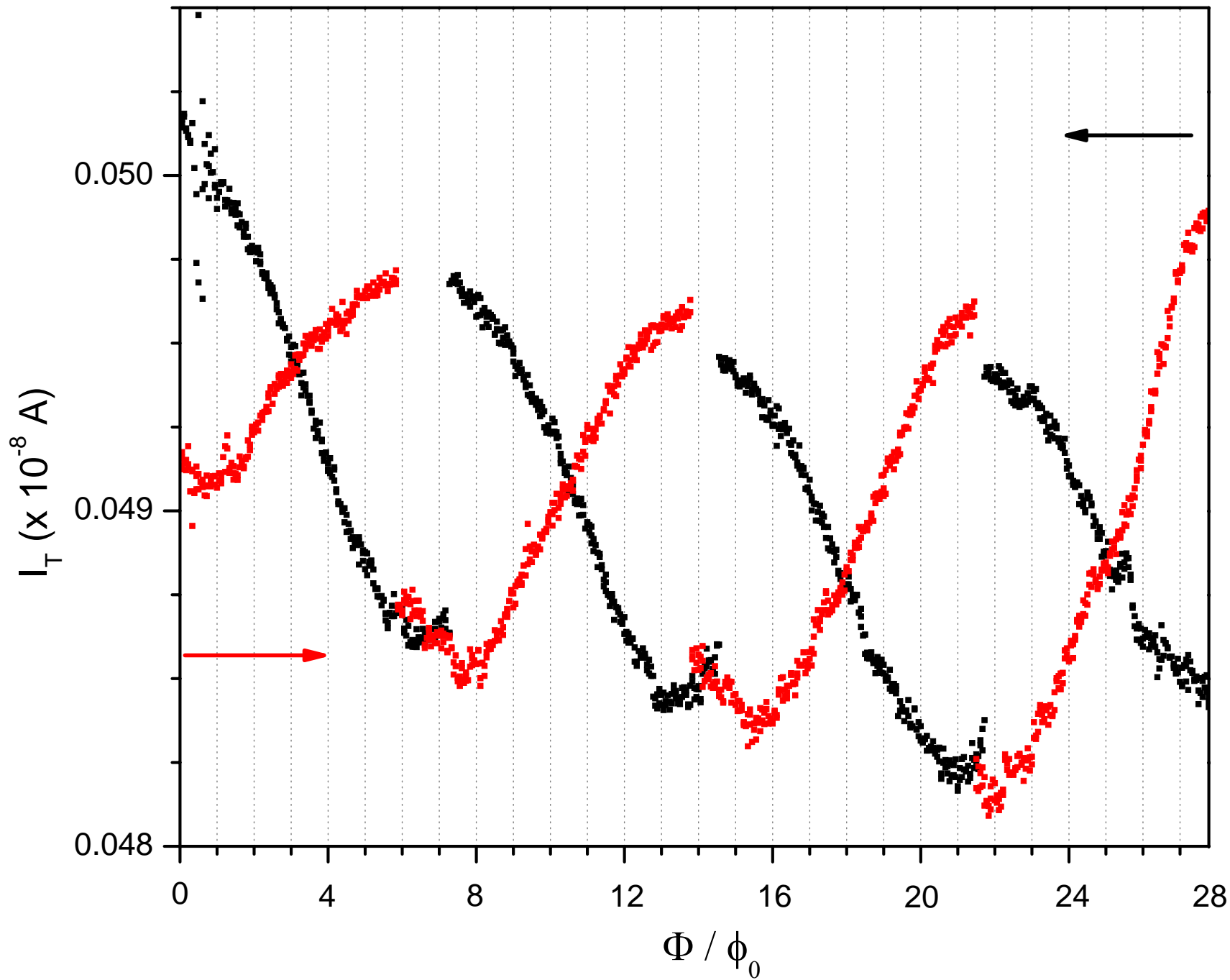

Supplementary figure 1. Tunnel current oscillations in a structure with uniform loop.

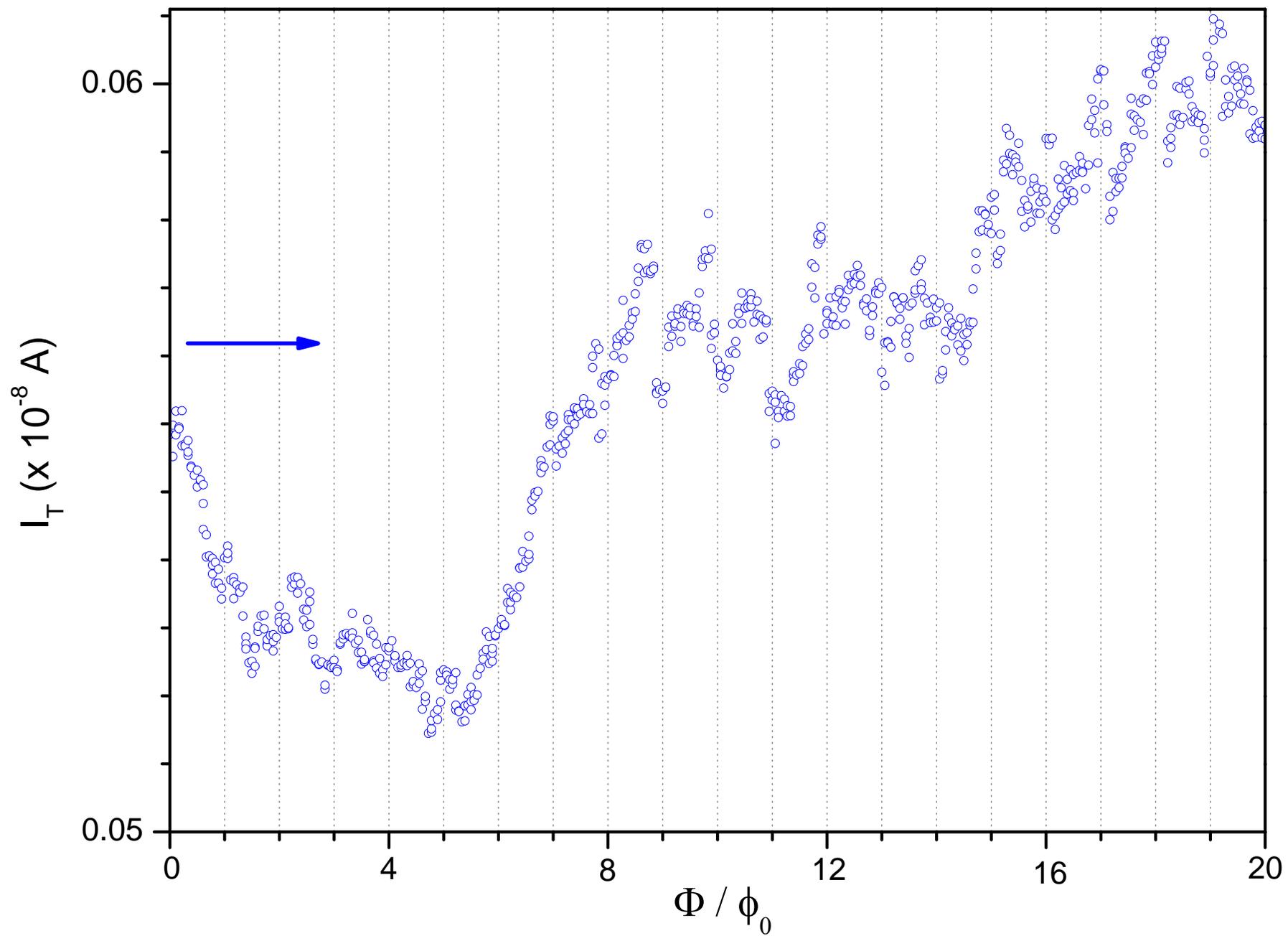

Supplementary figure 2. Tunnel current oscillations in a structure with the loop containing constriction